\documentclass{article}
\usepackage{amssymb}
\usepackage{latexsym}

\def\nqq{\hspace{-2em}}

\newcommand{\beq}[1]{\begin{equation}\label{#1}}
\newcommand{\eeq}{\end{equation}}
\newcommand{\bear}[1]{\begin{eqnarray}\label{#1}}
\newcommand{\ear}{\end{eqnarray}}
\newcommand{\nn}{\nonumber}

\catcode`\@=11 \@addtoreset{equation}{section}\catcode`\@=12

\def\barr{\left(\begin{array}}
\def\earr{\end{array}\right)}

\def\mm{\\ \nqq}

\newcommand{\N}{ {\mathbb N} }
\newcommand{\R}{ {\mathbb R} }

\newcommand{\sign}{ \mbox{\rm sign} }

\newcommand{\eps}{ \varepsilon }
\newcommand{\rank}{ \mbox{\rm rank} }

\newcommand{\p}{\partial}

\newcommand{\tri}{\Delta}

\newcommand{\fnm}{\footnotemark}
\newcommand{\fnt}{\footnotetext}

\begin{document}

\begin{center}

\large \bf Black brane solutions related to non-singular Kac-Moody
algebras

\end{center}

\vspace{0.3truecm}

\begin{center}

\normalsize\bf V. D. Ivashchuk\fnm[1]\fnt[1]{e-mail:
ivashchuk@mail.ru} and V. N. Melnikov\fnm[2]\fnt[2]{e-mail:
melnikov@phys.msu.ru},

\vspace{0.3truecm}

\it Center for Gravitation and Fundamental Metrology, VNIIMS,
Ozyornaya ul. 46, Moscow 119361, Russia

\it Institute of Gravitation and Cosmology, Peoples' Friendship
University of Russia,   Miklukho-Maklaya ul. 6, Moscow 117198,
Russia

\end{center}

\begin{abstract}

A multidimensional gravitational model containing scalar fields
and antisymmetric forms is considered. The manifold is chosen in
the form $M = M_0 \times M_1 \times \ldots \times M_n$, where
$M_i$ are Einstein spaces ($i \geq 1$). The sigma-model approach
and exact solutions with intersecting composite branes (e.g.,
solutions with harmonic functions and black brane ones) with
intersection rules related to non-singular Kac-Moody (KM) algebras
(e.g. hyperbolic ones) are considered.  Some examples  of black
brane solutions are  presented, e.g., those corresponding to
hyperbolic KM algebras: $H_2(q,q)$ ($q > 2$), $HA_2^{(1)} =
A_2^{++}$ and to the Lorentzian KM algebra  $P_{10}$.

\end{abstract}

\newpage

\section{\bf Introduction}
\setcounter{equation}{0}

In this paper, we consider certain classes of black brane
solutions related to non-singular   Kac-Moody  algebras. At
present, Kac-Moody (KM) Lie algebras \cite{Kac0,Moody} play
rather an important  role in different areas of mathematical
physics (see \cite{Kac,FS,Nik,HPS} and references therein).  We
recall that KM Lie algebra is a Lie algebra  generated by the
relations \cite{Kac}
 \bear{0.2}
[h_i,h_j] =0, \quad [e_i,f_j] = \delta_{ij} h_j, \mm \label{0.3}
[h_i,e_j] = A_{ij} e_j, \quad [h_i,f_j] = -A_{ij} f_j, \mm
\label{0.4} ({\rm ad} e_i)^{1- A_{ij}}(e_j) = 0 \quad (i \neq j),
\mm \label{6.5} ({\rm ad} f_i)^{1- A_{ij}}(f_j) = 0 \quad (i \neq
j). \ear

Here $A = (A_{ij})$ is a generalized  Cartan matrix,     $i,j = 1,
\ldots, r$, and   $r$ is the rank of the KM algebra. It means that
all  $A_{ii} =   2$;   $A_{ij}$ for $i \neq j$ are non-positive
integers and $A_{ij} =0 $ implies $A_{ji} =0$.    In what follows,
the matrix $A$ is restricted to    be non-degenerate (i.e. $\det A
\neq 0$) and symmetrizable,    i.e. $A = B {\cal D}$, where $B$ is
a symmetric matrix and    ${\cal D}$ is an invertible diagonal
matrix.

If $A$ is positive-definite (the Euclidean case) we get ordinary
finite-dimensional Lie algebras \cite{Kac,FS}. For non-Euclidean
signatures of $A$     all KM algebras are infinite-dimensional.
Among these the Lorentzian     KM algebras  with pseudo-Euclidean
signatures  $(-,+, \ldots, +)$  of the Cartan matrix $A$ are of
current interest  since they contain a subclass  of the so-called
hyperbolic KM algebras widely used in modern mathematical physics.
Hyperbolic KM algebras are by definition Lorentzian Kac-Moody
algebras with the property that, removing any node from their
Dynkin  diagram leaves one with a Dynkin diagram of the affine or
finite type.  The hyperbolic KM algebras were completely
classified in  \cite{Sac,BS}. They have  rank $2 \leq r \leq 10$.
For $r \geq 3$  there is a finite number of hyperbolic algebras.
For rank 10, there  are four algebras, known as $E_{10}$,
$BE_{10}$, $CE_{10}$  and $DE_{10}$. Hyperbolic KM algebras
appeared in ordinary gravity  \cite{FF} (${\cal F}_3 = AE_3 =
H_3$),  supergravity: \cite{J,Miz} ($E_{10}$),  \cite{Nic1}
(${\cal F}_3$), strings \cite{Moore}, oscillating behaviour near
the singularity \cite{DamH3} (see also \cite{DHN,IM-brev}), etc.

It has been  proposed by P. West that the Lorentzian
(non-hyperbolic) KM algebra $E_{11}$   is responsible for  a
hidden algebraic structure   characterizing  $11D$ supergravity
\cite{W}.  The same very extended algebra occurs in $IIA$ \cite{W}
and $IIB$ supergravities \cite{SW}.

Here we briefly consider another possibility    of utilizing
non-singular (e.g. hyperbolic) KM algebras,  suggested  in our
three papers \cite{IMBl,GrI,IKM}.    This possibility also
implicitly assumed in    \cite{IK,IMp1,IMp2,IMp3,IM-top} is
related to certain classes of  exact solutions describing
intersecting composite branes  in a multidimensional gravitational
model containing scalar fields  and antisymmetric forms  defined
on (warped) product  manifolds $M = M_0 \times M_1 \times \ldots
\times M_n$, where  $M_i$ are Ricci-flat spaces ($i \geq 1$). From
a pure mathematical point of view, these solutions  may be
obtained from sigma-models or Toda chains  corresponding to
certain non-singular KM algebras.  The information about a
(hidden) KM algebra is encoded in intersection rules which relate
the dimensions of brane intersections  with non-diagonal
components of the generalized Cartan matrix $A$ \cite{IMJ}. We
deal here with generalized Cartan matrices of the form
\beq{1.1}
A_{ss'} \equiv \frac{2(U^s,U^{s'})}{(U^{s'},U^{s'})},
\eeq
 $s,s'\in S$, with $(U^s,U^s)\neq 0$, for all $s\in S$ ($S$ is a
finite set). Here $U^s$ are the so-called brane (co-)vectors. They
are linear functions on $\R^N$, where $N = n+l$ and  $l$ is the
number of scalar fields. The indefinite   scalar product $(.,.)$
\cite{IMC} is defined on $(\R^N)^{*}$ and has the signature
$(-1,+1, \ldots, +1)$  if all scalar fields have positive kinetic
terms, i.e. there  are no phantoms (or ghosts).   The matrix   $A$
is symmetrizable.  $U^s$-vectors may be put in one-to-one
correspondence with simple roots  $\alpha_s$ of the generalized KM
algebra after a suitable normalization.

For   $D=11$ supergravity  \cite{CJS} and ten-dimensional  $IIA$,
$IIB$ supergravities all $(U^s,U^s) = 2$ \cite{IMJ,EH},  and
corresponding KM algebras are simply laced. It was shown in our
papers \cite{IMb2,IMb3} that the inequality $(U^s,U^s)> 0$ is a
necessary condition for the formation  of a billiard wall (if one
approaches the singularity) by the $s$-th  matter source (e.g., a
fluid component or a brane).

The scalar products  for brane vectors $U^s$  were found in
\cite{IMC}
\beq{1.2} (U^s,U^{s'})= d_{ss'}+\frac{d_s d_{s'}}{2-D}+
\chi_s \chi_{s'} <\lambda_{s}, \lambda_{s'}> , \eeq
where $d_s$
and $d_{s'}$ are the dimensions of  brane world volumes
corresponding to branes $s$ and $s'$, respectively, $d_{ss'}$ is
the dimension of the intersection of  brane world volumes, $D$ is
the total space-time dimension, $\chi_s = + 1, -1$ for electric or
magnetic brane respectively, and $<\lambda_{s}, \lambda_{s'}>$ is
the non-degenerate scalar product of the $l$-dimensional dilatonic
coupling vectors $\lambda_{s}$ and $\lambda_{s'}$  corresponding
to branes $s$ and $s'$.

The relations (\ref{1.1}), (\ref{1.2}) determine the brane
intersection rules \cite{IMJ}
\beq{1.3} d_{s s'}= d_{s s'}^{o} +
\frac12 K_{s'} A_{s s'},
\eeq
 $s \ne s'$, where $K_s = (U^s,U^s)$ and
\beq{1.4} d_{s s'}^{o} =
\frac{d_s d_{s'}}{D-2} -  \chi_s\chi_{s'} <\lambda_{s},
\lambda_{s'}>
\eeq
 is the dimension of the so-called orthogonal (or ($A_1 \oplus
A_1$)-) intersection of branes following from the orthogonality
condition  \cite{IMC}
\beq{1.5}
(U^s,U^{s'})= 0,
\eeq
 $s \ne s'$.

The relations (\ref{1.2}) and (\ref{1.4}) were derived in
\cite{IMC} under rather general assumptions: the branes were
composite, the number of scalar fields $l$ was arbitrary as well
as the signature of the bilinear form $<.,.>$ (or, equivalently,
the signature of the kinetic term for scalar fields),  Ricci-flat
factor spaces $M_i$ had arbitrary dimensions $d_i$ and signatures.
The intersection rules  (\ref{1.4}) appeared earlier in
\cite{AR,AEH} for all $d_i =1$ ($i > 0$) and $<.,.>$ being
positive-definite.

\section{\bf The model}
\setcounter{equation}{0}

\subsection{The action }

We consider a model governed by the action
\bear{2.1} S =&&
\frac{1}{2\kappa^{2}} \int_{M} d^{D}z \sqrt{|g|} \{ {R}[g] - 2
\Lambda - h_{\alpha\beta} g^{MN} \partial_{M} \varphi^\alpha
\partial_{N} \varphi^\beta
 \\ \nn
&& - \sum_{a \in {\bf \Delta}} \frac{\theta_a}{n_a!} \exp[ 2
\lambda_{a} (\varphi) ] (F^a)^2_g \}  + S_{GH}, \ear
 where $g = g_{MN} dz^{M} \otimes dz^{N}$ is the metric on the
manifold $M$, ${\dim M} = D$, $\varphi=(\varphi^\alpha)\in \R^l$
is a vector of dilatonic scalar fields, $(h_{\alpha \beta})$ is a
non-degenerate symmetric  $l\times l$ matrix ($l\in \N$),
$\theta_a  \neq 0$,
$$ F^a =  dA^a =  \frac{1}{n_a!} F^a_{M_1
\ldots M_{n_a}} dz^{M_1} \wedge \ldots \wedge dz^{M_{n_a}}$$
 is an $n_a$-form ($n_a \geq 2$) on the $D$-dimensional manifold $M$,
$\Lambda$ is a cosmological constant and $\lambda_{a}$ is a
$1$-form on $\R^l$: $\lambda_{a} (\varphi) =\lambda_{a
\alpha}\varphi^\alpha$, $a \in {\bf \Delta}$, $\alpha=1,\ldots,l$.
In (\ref{2.1}) we denote $|g| = |\det (g_{MN})|$, $(F^a)^2_g =
F^a_{M_1 \ldots M_{n_a}} F^a_{N_1 \ldots N_{n_a}} g^{M_1 N_1}
\ldots g^{M_{n_a} N_{n_a}}$, $a \in {\bf \Delta}$, where ${\bf
\Delta}$ is  some finite set (for example, of positive integers),
and $S_{\rm GH}$ is the standard Gibbons-Hawking boundary term
\cite{GH}. In models with one time all $\theta_a  =  1$  and the
signature of the metric is $(-1,+1, \ldots, +1)$;  $\kappa^{2}$ is
the multidimensional gravitational constant.

\subsection{The Ansatz for composite  branes }

Consider the manifold
\beq{2.10} M = M_{0}  \times M_{1} \times
\ldots \times M_{n}, \eeq
 with the metric
\beq{2.11} g= e^{2{\gamma}(x)} \hat{g}^0  + \sum_{i=1}^{n}
e^{2\phi^i(x)} \hat{g}^i , \eeq
 where $g^0  = g^0 _{\mu \nu}(x)
dx^{\mu} \otimes dx^{\nu}$ is an arbitrary metric with any
signature on the manifold $M_{0}$ and $g^i  = g^{i}_{m_{i}
n_{i}}(y_i) dy_i^{m_{i}} \otimes dy_i^{n_{i}}$ is a metric on
$M_{i}$  satisfying the equation
\beq{2.13}
 R_{m_{i}n_{i}}[g^i ] = \xi_{i} g^i_{m_{i}n_{i}},
\eeq
 $m_{i},n_{i}=1, \ldots, d_{i}$; $\xi_{i}= {\rm const}$,
$i=1,\ldots,n$. Here $\hat{g}^{i} = p_{i}^{*} g^{i}$ is the
pullback of the metric $g^{i}$  to the manifold  $M$ by the
canonical projection: $p_{i} : M \rightarrow  M_{i}$, $i =
0,\ldots, n$. Thus, $(M_i, g^i )$  are Einstein spaces, $i =
1,\ldots, n$. The functions $\gamma, \phi^{i} : M_0 \rightarrow \R
$ are smooth. We denote $d_{\nu} = {\rm dim} M_{\nu}$; $\nu = 0,
\ldots, n$;  $D = \sum_{\nu = 0}^{n} d_{\nu}$. We assume all
manifolds $M_{\nu}$, $\nu = 0,\ldots, n$, to be oriented and
connected. Then the volume $d_i$-form
\beq{2.14} \tau_i  \equiv
\sqrt{|g^i(y_i)|} \ dy_i^{1} \wedge \ldots \wedge dy_i^{d_i}, \eeq
 and signature parameter
\beq{2.15} \varepsilon(i)  \equiv {\rm
sign}( \det (g^i_{m_i n_i})) = \pm 1
\eeq
 are correctly defined for all $i=1,\ldots,n$.

Let $\Omega = \Omega(n)$  be a set of all non-empty subsets of $\{
1, \ldots,n \}$. The number of elements in $\Omega$ is $|\Omega| =
2^n - 1$. For any $I = \{ i_1, \ldots, i_k \} \in \Omega$, $i_1 <
\ldots < i_k$, we denote
\bear{2.16}
\tau(I) \equiv \hat{\tau}_{i_1}  \wedge \ldots \wedge \hat{\tau}_{i_k},  \\
\label{2.17}
\eps(I) \equiv \eps(i_1) \ldots \eps(i_k),  \\
\label{2.18}
M_{I} \equiv M_{i_1}  \times  \ldots \times M_{i_k}, \\
\label{2.19} d(I) \equiv  \sum_{i \in I} d_i. \ear

Here $\hat{\tau}_{i} = p_{i}^{*} \hat{\tau}_{i}$ is the pullback
of the form $\tau_i$  to the manifold  $M$ by the canonical
projection: $p_{i} : M \rightarrow  M_{i}$, $i = 1,\ldots, n$. We
also put $\tau(\emptyset)= \eps(\emptyset)= 1$ and
$d(\emptyset)=0$.

For fields of forms we consider the following composite
electromagnetic ansatz:
 \beq{2.1.1}
F^a=\sum_{I\in\Omega_{a,e}}{\cal F}^{(a,e,I)}+
\sum_{J\in\Omega_{a,m}}{\cal F}^{(a,m,J)}
\eeq
 where
\bear{2.1.2}
{\cal F}^{(a,e,I)}=d\Phi^{(a,e,I)}\wedge\tau(I), \\
\label{2.1.3} {\cal F}^{(a,m,J)}=
e^{-2\lambda_a(\varphi)}*(d\Phi^{(a,m,J)} \wedge\tau(J))
\ear
 are  elementary forms of electric and magnetic types, respectively,
$a \in {\bf \Delta}$, $I \in \Omega_{a,e}$, $J \in \Omega_{a,m}$
and  $\Omega_{a,v} \subset \Omega$, $v = e,m$. In (\ref{2.1.3})
$*=*[g]$ is the Hodge operator on $(M,g)$.

For scalar functions we put
\beq{2.1.5}
\varphi^\alpha=\varphi^\alpha(x), \quad \Phi^s=\Phi^s(x),
\eeq
 $s\in S$. Thus $\varphi^{\alpha}$ and $\Phi^s$ are functions on
$M_0$.

Here and below
\beq{2.1.6}
S=S_e \sqcup S_m, \quad S_v= \sqcup_{a
\in {\bf \Delta}} \{ a \} \times \{v \} \times \Omega_{a,v},
\eeq
 $v=e,m$. Here and in what follows $\sqcup$ means the union of
non-intersecting sets. The set $S$ consists of elements
$s=(a_s,v_s,I_s)$, where $a_s \in {\bf \Delta}$ is color index,
$v_s = e, m$ is the electro-magnetic index and the set $I_s \in
\Omega_{a_s,v_s}$ describes the location of a brane.

Due to (\ref{2.1.2}) and (\ref{2.1.3})
\beq{2.1.7}
d(I)=n_a-1,
\quad d(J)=D-n_a-1,
\eeq
 for  $I \in \Omega_{a,e}$ and $J \in \Omega_{a,m}$ (i.e., in the electric and
magnetic case, respectively).

\subsection{The sigma model}

Let $d_0 \neq 2$ and
\beq{2.2.1} \gamma=\gamma_0(\phi) \equiv
\frac1{2-d_0}\sum_{j=1}^nd_j\phi^j,
\eeq
 i.e., the generalized harmonic gauge (frame) is used.

We put two restrictions on the  sets of branes that guarantee the
block-diagonal form of the  energy-momentum tensor and the
existence of the sigma-model representation (without additional
constraints):

\beq{2.2.2a}
{\bf (R1)} \quad d(I \cap J) \leq d(I) - 2,
\eeq
 for any $I,J \in\Omega_{a,v}$, $a \in {\bf \Delta}$, $v= e,m$ (here
$d(I) = d(J)$) and

\beq{2.2.3a} {\bf (R2)} \quad d(I \cap J) \neq 0 \ for \ d_0 = 1,
\qquad d(I \cap J) \neq 1 \quad for \ d_0  = 3. \eeq

It was proved in \cite{IMC} that equations of motion for the model
(\ref{2.1}) and the Bianchi identities:

\beq{2.2.6} d{\cal F}^s=0,
\eeq
 $s \in S_m$, for fields from (\ref{2.11}),
(\ref{2.1.1})-(\ref{2.1.5}), under the restrictions (\ref{2.2.2a})
and (\ref{2.2.3a}), are equivalent to the equations of motion of
the $\sigma$-model governed by the action

\bear{2.2.7}
S_{\sigma 0} = && \frac{1}{2 \kappa_0^2} \int
d^{d_0}x\sqrt{|g^0|}\biggl\{R[g^0]-\hat G_{AB}
g^{0\mu\nu}\p_\mu\sigma^A\p_\nu\sigma^B  \\ \nn &&-\sum_{s\in
S}\eps_s \exp{(-2U_A^s\sigma^A)} g^{0\mu\nu}
\p_\mu\Phi^s\p_\nu\Phi^s - 2V \biggr\},
\ear
 where $(\sigma^A)=(\phi^i,\varphi^\alpha)$, $k_0 \neq 0$, the
index set  $S$ is defined in (\ref{2.1.6}),

\beq{2.2.8}
V = {V}(\phi)  = \Lambda e^{2 {\gamma_0}(\phi)}
-\frac{1}{2} \sum_{i =1}^{n} \xi_i d_i e^{-2 \phi^i + 2
{\gamma_0}(\phi)}
\eeq
 is the potential,
\beq{2.2.9}
(\hat G_{AB})= \barr{cc}
G_{ij}& 0\\
0& h_{\alpha\beta}
\earr \eeq
 is the target space metric with
 \beq{2.2.10}
 G_{ij}= d_i \delta_{ij}+\frac{d_i d_j}{d_0-2}
\eeq
and co-vectors
\beq{2.2.11}
U_A^s =   U_A^s \sigma^A =
\sum_{i \in I_s} d_i \phi^i - \chi_s \lambda_{a_s}(\varphi), \quad
(U_A^s) = (d_i \delta_{iI_s}, -\chi_s \lambda_{a_s \alpha}),
\eeq
 $s=(a_s,v_s,I_s)$. Here $\chi_e=+1$ and $\chi_m=-1$;

\beq{2.2.12}
\delta_{iI}=\sum_{j\in I}\delta_{ij}
\eeq
 is an indicator of $i$ belonging to $I$: $\delta_{iI}=1$ for $i\in
I$ and $\delta_{iI}=0$ otherwise; and

\bear{2.2.13}
\eps_s=(-\eps[g])^{(1-\chi_s)/2}\eps(I_s)
\theta_{a_s},
\ear
 $s\in S$, $\eps[g]\equiv\sign\det(g_{MN})$. More explicitly,
(\ref{2.2.13}) reads

\beq{2.2.13a}
\eps_s=\eps(I_s) \theta_{a_s} \ {\rm for} \ v_s = e;
\qquad \eps_s = -\eps[g] \eps(I_s) \theta_{a_s},
 \ {\rm for} \ v_s  = m.
\eeq

For finite internal space volumes $V_i$ (e.g. compact $M_i$) and
electric $p$-branes  (i.e. all $\Omega_{a,m} = \emptyset$) the
action (\ref{2.2.7}) coincides with the action (\ref{2.1}) when
$\kappa^{2} = \kappa^{2}_0 \prod_{i=1}^{n} V_i$.

\section{Solutions governed by harmonic functions}

\subsection{Solutions with a block-orthogonal
             set of $U^s$ and Ricci-flat factor-spaces}

Here we consider a special class of solutions to the equations of
motion governed by several harmonic functions, where all factor
spaces are Ricci-flat and the cosmological constant is zero, i.e.,
$\xi_i = \Lambda = 0$, $i = 1,\ldots,n$. In certain situations
these solutions describe extremal black branes charged by fields
of forms.

The solutions crucially depend upon  scalar products of
$U^s$-vectors $(U^s,U^{s'})$; $s,s' \in S$, where
\beq{3.1.1}
(U,U')=\hat G^{AB} U_A U'_B, \eeq for $U = (U_A), U' = (U'_A) \in
\R^N$, $N = n + l$ and \beq{3.1.2} (\hat G^{AB})
=\left(\begin{array}{cc}
 G^{ij}& 0\\
 0     & h^{\alpha\beta}
\end{array}\right)
\eeq
 is the inverse matrix  to  the matrix (\ref{2.2.9}). As
in \cite{IMZ}, we have
\beq{3.1.3}
G^{ij}=\frac{\delta^{ij}}{d_i}+\frac1{2-D},
\eeq
 $i,j=1,\dots,n$.

The scalar products (\ref{3.1.1}) of the vectors $U^s$  were
calculated in \cite{IMC} and are given by
\beq{3.1.4}
(U^s,U^{s'})=d(I_s\cap I_{s'})+\frac{d(I_s)d(I_{s'})}{2-D}+
\chi_s\chi_{s'}\lambda_{a_s \alpha} \lambda_{a_{s'} \beta}
h^{\alpha \beta}, \eeq
where
$(h^{\alpha\beta})=(h_{\alpha\beta})^{-1}$, and $s=(a_s,v_s,I_s)$,
$s'=(a_{s'},v_{s'},I_{s'})$ belong to $S$. This relation is  very
important one since it encodes  brane data (e.g., intersections)
via the  scalar products of $U$-vectors.

Let
\beq{3.1.5} S=S_1 \sqcup \dots \sqcup S_k,
\eeq
$S_i\ne\emptyset$, $i=1,\dots,k$, and
\beq{3.1.6} (U^s,U^{s'})=0
\eeq
for all $s\in S_i$, $s'\in S_j$, $i\ne j$; $i,j=1,\dots,k$.
The relation (\ref{3.1.5}) means that the set $S$ is a union of
$k$ non-intersecting (non-empty) subsets $S_1,\dots,S_k$.
According to (\ref{3.1.6}) the set of vectors $(U^s, s \in S)$ has
a block-orthogonal structure with respect to the scalar product
(\ref{3.1.1}), i.e., it  splits into $k$ mutually orthogonal
blocks $(U^s, s \in S_i)$, $i=1,\dots,k$.

Here we consider exact solutions in the model (\ref{2.1}) where
vectors $(U^s,s\in S)$ obey the block-orthogonal decomposition
(\ref{3.1.5}), (\ref{3.1.6}) with scalar products defined in
(\ref{3.1.4}) \cite{IMBl}. These solutions were obtained from the
corresponding solutions to the $\sigma$-model equations of motion
\cite{IMBl}.

{\bf Proposition  1.} {\em Let $(M_0,g^0)$ be Ricci-flat:
 $R_{\mu\nu}[g^0]=0$. Then the field configuration

\beq{3.1.7}
 g^0, \qquad \sigma^A=\sum_{s\in
S}\eps_sU^{sA}\nu_s^2\ln H_s, \qquad \Phi^s=\frac{\nu_s}{H_s},
\eeq
 $s\in S$,
 satisfies the field equations corresponding to the
action (\ref{2.2.7}) with $V=0$ if the real numbers $\nu_s$ obey
the relations

\beq{3.1.8}
\sum_{s'\in S}(U^s,U^{s'})\eps_{s'}\nu_{s'}^2=-1
\eeq
 $s\in S$, the functions $H_s >0$ are harmonic, i.e.
$\tri[g^0]H_s=0$, $s\in S$, and $H_s$  coincide inside the blocks:
$H_s=H_{s'}$ for $s,s'\in S_i$, $i=1,\dots,k$.}

Using the sigma-model solution from Proposition 1 and the
relations for contra-variant components \cite{IMC}:

\beq{2.2.38} U^{si}=\delta_{iI_s}-\frac{d(I_s)}{D-2}, \quad
 U^{s\alpha}=-\chi_s\lambda_{a_s}^\alpha,
\eeq
 $s=(a_s,v_s,I_s)$,
 we get \cite{IMBl}:
\bear{3.1.11} g= \left(\prod_{s \in S}
H_s^{2d(I_s)\eps_s\nu_s^2}\right)^{1/(2-D)} \left\{\hat{g}^0+
\sum_{i=1}^n \left(\prod_{s \in S}
H_s^{2\eps_s\nu_s^2\delta_{iI_s}}\right)
\hat{g}^i\right\}, \\
\label{3.1.14}
\varphi^\alpha= -\sum_{s \in S}\lambda_{a_s}^\alpha
\chi_s
\eps_s\nu_s^2\ln H_s, \\
\label{3.1.15}
F^a=\sum_{s\in S}{\cal F}^s\delta_{a_s}^a,
\ear
 where $i=1,\dots,n$, $\alpha=1,\dots,l$, $a \in {\bf \Delta}$ and

\bear{3.1.16}
{\cal F}^s=\nu_s dH_s^{-1}\wedge\tau(I_s), \mbox{
for } v_s=e, \mm \label{3.1.17} {\cal F}^s=\nu_s
(*_0dH_s)\wedge\tau(\bar I_s), \mbox{ for } v_s=m,
\ear
 $H_s$ are harmonic functions on $(M_0,g^0)$ which coincide inside
the blocks (i.e., $H_s=H_{s'}$ for $s,s'\in S_i$, $i=1,\dots,k$)
and the relations  (\ref{3.1.8}) on the parameters $\nu_s$ are
imposed. The matrix $((U^s,U^{s'}))$ and the parameters $\eps_s$,
$s\in S$, are defined in (\ref{3.1.4}) and (\ref{2.2.13}),
respectively; $\lambda_a^\alpha=  h^{\alpha\beta}\lambda_{a
\beta}$, $*_0=*[g^0]$ is the Hodge operator  on $(M_0,g^0)$ and

\beq{2.2.5I} \bar I = \{1, \dots, n \} \setminus I \eeq is the
dual set.  (In (\ref{3.1.17}) we have redefined the sign of the
parameter $\nu_{s}$.)

\subsection{Solutions related to non-singular KM algebras}

Now we will study the  solutions (\ref{3.1.11})-(\ref{3.1.17}) in
more detail and show that some of them may be related to
non-singular KM algebras. We put
\beq{3.1.2.1}
  K_s \equiv (U^s,U^s)\neq 0
\eeq
 for all $s\in S$ and introduce the  matrix $A=(A_{ss'})$:
\beq{3.1.2.2}
 A_{ss'} \equiv \frac{2(U^s,U^{s'})}{(U^{s'},U^{s'})},
 \eeq
$s,s'\in S$. Here  some ordering in $S$ is assumed.

Using this definition and (\ref{3.1.4}) we obtain the intersection
rules \cite{IMJ}

\beq{3.1.2.3} d(I_{s}\cap I_{s'})=\Delta(s,s')+\frac12 K_{s'} A_{s
s'}, \eeq where $s \ne s'$, and \beq{3.1.D} \Delta(s,s') =
\frac{d(I_s)d(I_{s'})}{D-2} - \chi_s\chi_{s'}\lambda_{a_s \alpha}
\lambda_{a_{s'} \beta} h^{\alpha \beta}
\eeq
 defines the so-called ``orthogonal'' intersection rules \cite{IMC}
(see also  \cite{AR,AEH} for $d_i = 1$).

In $D = 11$ and $D = 10$ ($IIA$ and $IIB$ ) supergravity models,
all $K_s = 2$ and hence the intersection rules (\ref{3.1.2.3}) in
this case have a simpler form \cite{IMJ}:
\beq{3.1.2.3.sl}
d(I_{s}\cap I_{s'})= \Delta(s,s')+ A_{s s'}, \eeq
 where $s \ne s'$, implying $A_{s s'} = A_{s' s}$. The
corresponding KM algebra is simply-laced in this case.

For $\det A \neq 0$ relation (\ref{3.1.8}) may be rewritten in the
equivalent form
\beq{3.1.2.5} - \eps_s\nu_s^2(U^s,U^s)= 2
\sum_{s'\in S} A^{ss'} \equiv b_s,
\eeq
 where $s\in S$, and $(A^{ss'})=A^{-1}$. Thus eq. (\ref{3.1.8})
may be resolved in terms of $\nu_s$ for certain $\eps_s=\pm1$,
$s\in S$. We note that due to $(\ref{3.1.6})$ the matrix $A$ has a
block-diagonal structure  and, hence, for any $i$-th block the set
of parameters $(\nu_s,  s \in S_i)$ depends on the matrix inverse
to the matrix  $(A_{s s'};  s,  s' \in S_i)$.

Now  we consider the one-block case such that the brane
intersections are related to some  non-singular KM algebras.

{\bf  Finite-dimensional Lie algebras \cite{GrI}}

Let $A$ be the Cartan matrix of a simple finite-dimensional Lie
algebra. In this case $A_{ss'}\in\{0,-1,-2,-3\}$, $s\ne s'$. The
elements of the inverse matrix $A^{-1}$ are positive (see Ch. 7 in
\cite{FS}) and hence we get from (\ref{3.1.2.5}) the same
signature relation  as in the  orthogonal case \cite{IMC}:
\beq{3.1.2.20b} \eps_s(U^s,U^s) < 0, \eeq
 $s\in S$.

If all $(U^s,U^{s}) > 0$, we get  $\eps_{s} < 0$,  $s \in S$.
Moreover, all $b_s$ are positive integers: \beq{3.1.2.6}
 b_s = n_s \in \N,
\eeq
 $s \in S$.
The integers $n_s$ coincide with the components of the twice dual
Weyl vector in the basis of simple co-roots (see Ch. 3.1.7 in
(\cite{FS}).

{\bf  Hyperbolic KM algebras}

Let $A$ be a generalized Cartan matrix corresponding to a simple
hyperbolic KM algebra. For   hyperbolic algebras, the following
relations are satisfied
\beq{3.1.2.20}
\eps_s(U^s,U^s) >0, \eeq
 since all $b_s$ are negative in the hyperbolic case \cite{GOW}:
\beq{3.1.2.6b}
b_s  < 0,
\eeq
 where $s\in S$.

For $(U^s,U^{s}) > 0$ we get  $\eps_{s} > 0$, $s \in S$. If
$\theta_{a_s} > 0$ for all $s \in S$, then
\beq{3.1.28a}
\eps(I_s)
= 1   \ {\rm for} \ v_s = e; \qquad \eps(I_s) = - \eps[g] \ {\rm
for} \ v_s = m.
\eeq

For a pseudo-Euclidean metric $g$ all $\eps(I_s) = 1$, and hence
all branes are Euclidean or should contain even number of time
directions: $2,4, \ldots$. For $\eps[g] = 1$ only magnetic branes
may be pseudo-Euclidean.

{\bf $B_D$-models.} The $B_D$-model has the  action \cite{IMJ}
\beq{3.1.2.09}
 S_D=\int
 d^Dz\sqrt{|g|}\biggl\{R[g]+ g^{MN}\p_M\vec\varphi\p_N\vec\varphi-
\sum_{a=4}^{D-7}\frac1{a!}\exp[2\vec\lambda_a\vec\varphi](F^a)^2\biggr\},
\eeq
 where   $\vec\varphi=(\varphi^1,\dots,\varphi^l)\in\R^l$,
$\vec\lambda_a= (\lambda_{a1},\dots,\lambda_{al})\in\R^l$,
$l=D-11$, $\rank F^a=a$, $a=4,\dots,D-7$. Here vectors
$\vec\lambda_a$ satisfy the relations

\bear{3.1.2.010}
\vec\lambda_a\vec\lambda_b=N(a,b)-\frac{(a-1)(b-1)}{D-2} =
\Lambda_{ab}, \\
\label{3.1.2.011} N(a,b)=\min(a,b)-3, \ear $a,b=4,\dots,D-7$ and
$\vec\lambda_{D-7}=-2\vec\lambda_4$. For $D>11$ vectors
$\vec\lambda_4,\dots,\vec\lambda_{D-8}$ are linearly independent.
(It may be verified that the matrix $(\Lambda_{ab})$ is
positive-definite  and hence the set of vectors obeying
(\ref{3.1.2.010}) does exist.)

The model (\ref{3.1.2.09}) contains $l$ scalar fields with a
negative kinetic term (i.e.,
$h_{\alpha\beta}=-\delta_{\alpha\beta}$ in (\ref{2.1})) coupled to
several forms (the number of forms is $(l+1)$) .

For the brane worldvolumes we have the following dimensions (see
(\ref{2.1.7})):   $d(I)=3,\dots,D-8$ for $I\in\Omega_{a,e}$
 and $d(I)=D-5,\dots,6$ for $I \in\Omega_{a,m}$.
Thus there are $(l+1)$ electric and $(l+1)$ magnetic $p$-branes,
$p=d(I)-1$. In $B_D$-model all $K_s = 2$. (For $B_{12}$-model see
\cite{KKLP}.)

{\bf Example 1: $H_2(q_1,q_2)$ algebra \cite{IMBl}.} Let

\beq{3.13}
 A =\barr{cc}
 2& -q_1\\
 -q_2& 2 \earr, \quad q_1q_2>4,
\eeq
 $q_1,q_2\in \N$. This is the Cartan matrix of the hyperbolic KM
algebra $H_{2}(q_1,q_2)$ \cite{Kac}.  From (\ref{3.1.2.5}) we get
\beq{3.14} \eps_s\nu_s^2(U^s,U^s)(q_1q_2-4)=2q_s+4, \eeq
 $s\in\{1,2\}=S$.

An example of the $H_{2}(q,q)$-solution for $B_D$-model  with two
electric $p$-branes  ($p=d_1,d_2$), corresponding to $F^a$ and
$F^b$ fields and intersecting on a time manifold, is as follows
\cite{IMBl}: \bear{3.1.2.33}
g=H^{-2/(q-2)}\hat{g}^0-H^{2/(q-2)}dt\otimes
dt+\hat{g}^1+\hat{g}^2, \mm \label{3.1.2.34a}
F^a=\nu_{1}dH^{-1}\wedge dt\wedge \hat{\tau}_1, \mm
\label{3.1.2.34b} F^b = \nu_{2}dH^{-1}\wedge dt\wedge
\hat{\tau}_2, \mm \label{3.1.2.35}
\vec\varphi=-(\vec\lambda_a+\vec\lambda_b)(q -2)^{-1}\ln H \ear
 where $d_0=3$, $d_1=a - 2$, $a=q+4$, $a < b$, $d_2=b-2$, $D=a+b$,
and $\nu_{1}^2 = \nu_{2}^2 = (q - 2)^{-1}$. The signature
restrictions are : $\eps_1= \eps_2 = -1$. Thus the space-time
$(M,g)$ should contain at least three time directions. The minimal
$D$ is 15. For $D=15$ we get $a =7$, $b =8$, $d_1 =5$, $d_2= 6$
and $q=3$.

\section{Black brane solutions}

In this section we consider spherically symmetric solutions with
$d_0 =1$ and  $M_1 = S^{d_1}$, $g^1 = d \Omega^2_{d_1}$, where $d
\Omega^2_{d_1}$ is the canonical metric on a unit sphere
$S^{d_1}$, $d_1 \geq 2$. The manifold $M_0$ corresponds to a
radial variable $R$. We put also  $M_2 = \R$, $g^2 = - dt \otimes
dt$, i.e.,  $M_2$ is a time manifold and
\beq{5.2.18} 2 \in I_s,
\quad \forall s \in S,
\eeq
 i.e., all branes have a common time direction $t$.

In \cite{IMp1,IMp2,IMp3}, the following solutions with a horizon
were obtained:

\bear{5.2.30}
g= \Bigl(\prod_{s \in S}
 H_s^{2 h_s d(I_s)/(D-2)} \Bigr) \biggl\{ \left(1 -
\frac{2\mu}{R^{\bar{d}}}\right)^{-1} dR \otimes dR + R^2  d
\Omega^2_{d_1}  \\ \nn -  \Bigl(\prod_{s \in S} H_s^{-2 h_s}
\Bigr) \left(1 - \frac{2\mu}{R^{\bar{d}}}\right) dt \otimes dt +
\sum_{i = 3}^{n} \Bigl(\prod_{s \in S} H_s^{-2 h_s \delta_{iI_s}}
\Bigr) \hat{g}^i  \biggr\},
\\  \label{5.2.31}
\exp(\varphi^\alpha)= \prod_{s\in S} H_s^{h_s \chi_s
\lambda_{a_s}^\alpha}, \ear
 where $F^a= \sum_{s \in S} \delta^a_{a_s} {\cal F}^{s}$, and
\beq{5.2.32}
{\cal F}^s= - \frac{Q_s}{R^{d_1}} \left( \prod_{s'
\in S}  H_{s'}^{- A_{s s'}} \right) dR \wedge\tau(I_s), \eeq $s\in
S_e$,

\beq{5.2.33}
{\cal F}^s= Q_s \tau(\bar I_s),
\eeq $s \in S_m$.

Here $Q_s \neq 0$, $h_s =K_s^{-1}$, $s \in S$, and the generalized
Cartan matrix $(A_{s s'})$ is non-degenerate.

The functions $H_s > 0$ obey the equations
\beq{5.3.1}
\frac{d}{dz} \left( \frac{(1 - 2\mu z)}{H_s} \frac{d}{dz} H_s
\right) = \bar B_s \prod_{s' \in S}  H_{s'}^{- A_{s s'}}, \eeq

\bear{5.3.2a}
H_{s}((2\mu)^{-1} -0) = H_{s0} \in (0, + \infty), \\
\label{5.3.2b} H_{s}(+ 0) = 1, \ear
 $s \in S$, where $H_s(z) > 0$, $\mu > 0$, $z = R^{-\bar d} \in (0,
(2\mu)^{-1})$, $\bar d = d_1 -1$ and $\bar B_s = \eps_s K_s Q_s^2/
\bar d^2 \neq 0$.

There exist solutions to eqs. (\ref{5.3.1})-(\ref{5.3.2a}) of
polynomial type. The simplest example occurs in orthogonal case
\cite{BIM,IMJ} (for $d_i = 1$  see also \cite{AIV,Oh}):
$(U^s,U^{s'})= 0$, for  $s \neq s'$, $s, s' \in S$. In this case
$(A_{s s'}) = {\rm diag}(2,\ldots,2)$ is a Cartan matrix of the
semisimple Lie algebra ${\bf A_1} \oplus  \ldots  \oplus  {\bf
A_1}$ and

\beq{5.3.5}
H_{s}(z) = 1 + P_s z
\eeq
 with $P_s \neq  0$, satisfying
\beq{5.3.5a}
P_s(P_s + 2\mu) = -\bar B_s,
\eeq
 $s \in S$.

In \cite{Br1,IMJ2} this solution was generalized to the
block-orthogonal case  (\ref{3.1.5}), (\ref{3.1.6}). In this case
(\ref{5.3.5}) is modified as follows
 \beq{5.3.8} H_{s}(z) = (1 + P_s z)^{b_s}, \eeq
 where $b_s$ are defined as follows
\beq{5.2.20}
 b_s = 2 \sum_{s'
\in S} A^{s s'}
\eeq
 and parameters $P_s$ coincide within blocks, i.e., $P_s = P_{s'}$
for $s, s' \in S_i$, $i =1,\dots,k$. The parameters $P_s \neq 0 $
satisfy the relations \cite{IMJ2,IM-top}
\beq{5.3.5b}
P_s(P_s +
 2\mu) = -\bar B_s/b_s,
\eeq
 $s \in S$, and the parameters $\bar B_s/b_s$  coincide within
blocks, i.e. $\bar B_s/b_s = \bar B_{s'}/b_{s'}$ for $s, s' \in
S_i$, $i =1,\dots,k$.

{\bf Finite-dimensional Lie algebras.}

Let $(A_{s s'})$ be  a Cartan matrix  for a  finite-dimensional
semisimple Lie  algebra $\cal G$. In this case all powers in
(\ref{5.2.20})  are  positive integers  which coincide with the
components of twice the  dual Weyl vector in the basis of simple
co-roots \cite{FS}, and  hence all functions $H_s$ are
polynomials, $s \in S$.

{\bf Conjecture 1.} {\em Let $(A_{s s'})$ be  a Cartan matrix for
a semisimple finite-dimensional Lie algebra $\cal G$. Then  the
solutions to eqs. (\ref{5.3.1})-(\ref{5.3.2b}) (if any) have a
polynomial structure:
 \beq{5.3.12} H_{s}(z) = 1 + \sum_{k =
                1}^{n_s} P_s^{(k)} z^k,
 \eeq
 where $P_s^{(k)}$ are constants, $k = 1,\ldots, n_s$; $n_s = b_s =
2 \sum_{s' \in S} A^{s s'} \in \N$ and $P_s^{(n_s)} \neq 0$,  $s
\in S$}.

In the extremal case ($\mu = + 0$), an analogue of this conjecture
was suggested previously in \cite{LMMP}. Conjecture 1 was verified
for the  ${\bf A_m}$ and ${\bf C_{m+1}}$ Lie algebras in
\cite{IMp2,IMp3}. Explicit expressions for polynomials
corresponding to the Lie algebras $C_2$ and $A_3$ were obtained in
\cite{GrIK} and \cite{GrIM}, respectively. Recently, a family of
black brane type solutions in the model  with multicomponent
anisotropic fluid were found in  \cite{I-2010}.

{\bf Hyperbolic KM algebras.} Let $(A_{s s'})$ be  a Cartan matrix
for the infinite-dimensional hyperbolic KM  algebra $\cal G$. In
this case, all powers in (\ref{5.2.20})  are  negative,  and hence
we have no chance to get a polynomial structure for $H_s$. Here we
are led to an open problem of seeking  solutions to the set of
``master'' equations (\ref{5.3.1})-(\ref{5.3.2a}). These solutions
determine special solutions to the Toda-chain equations
corresponding to the hyperbolic KM algebra $\cal G$.

{\bf Example 2. Black hole solutions for the KM algebras  $A_1
 \oplus A_1$, $A_2$ and  $H_2(q,q)$  \cite{IM-SIGMA}.}

Consider a 4-dimensional model governed by the action
\bear{5.4.1}
S =  \int_{M} d^{4}z \sqrt{|g|} \{ {R}[g] - \eps g^{MN}
\partial_{M} \varphi\partial_{N} \varphi -  \frac{1}{2}
e^{2\lambda \varphi} (F^1)^2 -  \frac{1}{2} e^{- 2\lambda \varphi}
(F^2)^2 \}.
\ear

Here $F^1$ and $F^2$ are 2-forms, $\varphi$ is a scalar field and
$\eps = \pm 1$.

We consider a black brane solution   defined on $\R_{*} \times
S^{2} \times \R $   with two electric branes   $s_1$ and $s_2$,
corresponding to the forms $F^1$ and $F^2$,   respectively, with
the sets $I_1 = I_2 = \{ 2 \}$.   Here $\R_{*}$ is a subset of
$\R$, $M_1 = S^{2}$,  $g^1 = d \Omega^2_{2}$  is the canonical
metric on $S^{2}$, $M_2 = \R$, $g^2 = - dt \otimes dt$ and $\eps_1
= \eps_2 = -1$.

The scalar products of the $U$-vectors are (we identify $U^i =
U^{s_i}$):

\bear{5.4.2} (U^1,U^1) = (U^2, U^2) = \frac{1}{2 } + \eps
\lambda^2  \neq 0,    \qquad (U^1,U^2) = \frac{1}{2 } - \eps
\lambda^2.  \ear

The matrix $A$ from (\ref{3.1.2.2}) is a generalized
non-degenerate Cartan matrix  if and only if
\beq{5.4.3}
\frac{2(U^1,U^2)}{(U^2, U^2)} = -q,
\eeq
 or, equivalently,
\beq{5.4.4}
\eps \lambda^2 = \frac{2+q}{2(2-q)},
\eeq
 where $q = 0,1,3,4, ...$. This takes place if $\eps = + 1$ for
$q = 0, 1$    and   $\eps = - 1$ for $q =  3, 4, 5, \dots$.

The first branch ($\eps = + 1$) corresponds to the
finite-dimensional Lie algebras $A_1 \oplus A_1$ ($q =0$), $A_2$
($q = 1$) and  the second one ($\eps = - 1$)  to the hyperbolic KM
algebras $H_2(q,q)$, $q = 3,4, ...$. In the hyperbolic case, the
scalar field $\varphi$ is a phantom.

The  black brane solution reads (see \ref{5.2.30})-(\ref{5.2.32}))

\bear{5.4.7}
g= (H_1 H_2)^{h} \biggl\{ \left(1 - \frac{2\mu}{R}
\right)^{-1} dR \otimes dR + R^2  d \Omega^2_{2}
\\ \nonumber
 - (H_1 H_2)^{- 2h}
\left(1 - \frac{2\mu}{R}\right) dt \otimes dt \biggr\},
\\  \label{5.4.8}
\exp(\varphi)= (H_1 / H_2)^{\eps \lambda h},
\\  \label{5.4.9}
F^s= \frac{Q_s}{R^2} H_s^{-2} (H_{\bar s})^{q} dt \wedge dR,
\ear
 $s = 1, 2$. Here $h = (2- q)/2$ and $\bar s = 2, 1 $ for $s = 1,
2$, respectively.

The moduli functions $H_s > 0$ obey  the equations (see
(\ref{5.3.1}))
\beq{5.4.10}
\frac{d}{dz} \left( \frac{(1 - 2\mu
z)}{H_s} \frac{d}{dz} H_s \right) = \frac{2 Q_s^2}{q-2} H_s^{-2}
(H_{\bar s})^{q},
\eeq
 with the boundary conditions  $H_{s}((2\mu)^{-1} -0) = H_{s0} \in
(0, + \infty)$,   $H_{s}(+ 0) = 1$, $s = 1, 2$. Here $\mu
 > 0$, $z = 1/R \in (0, (2\mu)^{-1})$. For $q = 0, 1$ the solutions
to eqs. (\ref{5.4.10}) with these boundary conditions  were given
in \cite{IMp1,IMp2,IMp3}. They are polynomials of degrees $1$ and
$2$ for $q=0$ and $q =1$, respectively. For $q = 3, 4, ...$ the
exact solutions to eqs. (\ref{5.4.10}) are yet unknown.

In the special case $Q_1^2 = Q_2^2$ the metric coincides with that
 of the Reissner-Nordstr\"om solution \cite{IM-SIGMA}.

{\bf Example 3: Black brane solution corresponding to the KM
algebra $HA_{2}^{(1)} = A_2^{++}$.}

Now we consider the $B_{15}$-model in  $15$-dimensional
pseudo-Euclidean space of signature $(-, +, ..., +)$ with the
forms $F^4$,..., $F^8$.

Here we deal with four electric branes   $s_1, s_2, s_3, s_4$
corresponding to the 6-form $F^6$.   The brane sets are: $I_1 = \{
1, 2, 3,  11, 12  \}$, $I_2 = \{ 4, 5, 6, 11, 12 \}$, $I_3 = \{ 7,
8, 9,  11, 12  \}$, $I_4 = \{ 1, 4, 10, 11, 12 \}$.

It may be verified that these sets obey the intersection rules
corresponding to  the hyperbolic KM algebra $HA_{2}^{(1)}$ with
the Cartan matrix
\bear{5a.H.A.2}
   A=\barr{cccc}
    2 & -1  & -1 &  0  \\
   -1 &  2  & -1 &  0  \\
   -1 & -1  &  2 & -1  \\
    0 &  0  & -1 &  2  \\
 \earr,
\ear
(see (\ref{3.1.2.3.sl}) with $I_{s_i} = I_i$).

Now we give a black brane solution  for the configuration of four
branes under consideration.

In what follows the relations  $\eps_s = +1$ and $h_s = 1/2$, $s
\in S$, are used.

 The metric (\ref{5.2.30}) reads
\bear{5a.3.ha.g}
g= (H_1 H_2 H_{3} H_4)^{5/13}
   \biggl\{ \left(1 -  \frac{2\mu}{R} \right)^{-1} dR \otimes dR + R^2  d \Omega^2_{2}
\\ \nonumber
 -  (H_1 H_2 H_{3} H_4)^{-1}
 \left[ \left(1 -  \frac{2\mu}{R} \right) dt^1 \otimes dt^1  + dt^2 \otimes dt^2 \right]
\\ \nonumber
 + (H_1 H_4)^{-1}  dx^1 \otimes dx^1
 + H_1^{-1} [ dx^2 \otimes dx^2   +  dx^3 \otimes dx^3 ]
\\ \nonumber
 + (H_2 H_4)^{-1}  dx^4 \otimes dx^4
 +  H_2^{-1} [ dx^5 \otimes dx^5
 +   dx^6 \otimes dx^6 ]
\\ \nonumber
 + H_3^{-1} [ dx^7 \otimes  dx^7
 +  dx^8 \otimes dx^8   +  dx^9 \otimes  dx^9 ]
\\ \nonumber
 + H_4^{-1}  dx^{10} \otimes dx^{10}
\biggr\}.
\ear

Here $t^1 = x^{11}$ and  $t^2 = x^{12}$ are time-like variables

The non-zero form field is
 \bear{5a.3.ha.f} F^6 =
 - Q_{1} R^{-2} H_1^{-2} H_2 H_3 dR \wedge dt^1 \wedge dt^2 \wedge dx^{1} \wedge dx^{2}\wedge dx^{3}
\\ \nonumber
 - Q_{2} R^{-2} H_1 H_2^{-2} H_3 dR \wedge dt^1 \wedge dt^2 \wedge dx^{4} \wedge dx^{5}\wedge dx^{6}
\\ \nonumber
 - Q_{3} R^{-2} H_1 H_2 H_3^{-2} H_4  dR \wedge dt^1 \wedge dt^2 \wedge dx^{7} \wedge dx^{8}\wedge dx^{9}
\\ \nonumber
 - Q_{4} R^{-2} H_3 H_4^{-2} dR \wedge dt^1 \wedge dt^2 \wedge dx^{1} \wedge dx^{4}\wedge dx^{10},
\ear
 where $Q_s \neq 0$, $s = 1, 2, 3, 4$.

 The scalar fields  read
\beq{5a.3.ha.p}
\varphi^{\alpha} =
 - \frac{1}{2} \lambda_{6 \alpha} \ln(H_1 H_2 H_{3} H_4),
\eeq
 $\alpha = 1,2,3,4$. Here   $H_s > 0$ obey the equations
 \beq{5a.3.1}
 \frac{d}{dz} \left( \frac{(1 - 2\mu z)}{H_s}
 \frac{d}{dz} H_s \right) = 2 Q_s^2
 \prod_{s' = 1}^4  H_{s'}^{- A_{s s'}}, \eeq
 with the boundary conditions
 $H_{s}((2\mu)^{-1} -0) = H_{s0} \in
(0, + \infty)$, and  $H_{s}(+ 0) = 1$, $s = 1, ..., 4$. Here
$\mu > 0$, $z = R^{-1} \in (0, (2\mu)^{-1})$  and   $(A_{s s'})$ is
the Cartan matrix (\ref{5a.H.A.2}) for the KM algebra
 $HA_2^{(1)}$.

{\bf Special 1-block solution.}
 This solution is valid when  a
special set of charges is considered:
\beq{4a.3.ha.Q}
 Q_s^2 = Q^2 |b_s|,
\eeq
 where $Q \neq 0$ and
\beq{4a.3.ha.b}
 b_s = 2 \sum_{s' =1}^{4} A^{ss'} = -12, -12, -14, - 6,
\eeq
for $s = 1, 2, 3, 4$, respectively. In this case the moduli
functions read \beq{5a.4.11}
 H_s  = H^{b_s}, \qquad  H =  1 + P/R,
\eeq

where $P(P + 2\mu) = 2Q^2$. These functions obey $H > 0$ for $R
\in [2 \mu, + \infty)$     if $P > - 2 \mu$ ($\mu > 0$). Due to
this inequality and the     relation  $P(P + 2 \mu) = Q^2$,
 we get
 \beq{5a.4.12}  P = -  \mu + \sqrt{\mu^2 + 2Q^2} > 0. \eeq

The Hawking temperature in this case is \cite{IM-top} $T_H = (1 +
P/2\mu)^{22}/(8 \pi  \mu)$. It diverges  as $\mu \to + 0$. This is
in agreement with the fact that the metric (\ref{5a.3.ha.g}) has a
singularity at $R = + 0$ if  $\mu = + 0$.

{\bf Example 4: Black brane solution corresponding to the
Lorentzian KM algebra  $P_{10}$.}

Now we consider another solution for  the $B_{15}$-model in
$15$-dimensional pseudo-Euclidean space of signature $(-, +, ...,
+)$   with the non-zero 6-form $F^6$.

Here we deal with ten electric branes   $s_1, ..., s_{10}$
corresponding to the 4-form $F^4$.   The brane sets are:
 $I_1 = \{
1, 4, 7, 11, 12  \}$, $I_2 = \{ 8, 9, 10, 11, 12 \}$, $I_3 = \{ 2,
5, 7, 11, 12  \}$, $I_4 = \{ 4, 6, 10, 11, 12 \}$, $I_5 = \{ 2, 3,
9, 11, 12  \}$, $I_6 = \{ 1, 2, 8, 11, 12 \}$, $I_7 = \{ 1, 3, 10,
11, 12 \}$, $I_8 = \{ 4, 5, 8, 11, 12 \}$, $I_9 = \{ 3, 6, 7, 11,
12  \}$, $I_{10} = \{ 5, 6, 9, 11, 12 \}$.

These sets obey  the intersection rules corresponding to the
Lorentzian KM algebra $P_{10}$ (we call it the Petersen algebra)
with the following Cartan matrix
\bear{5a.P.A.2}
A=\barr{cccccccccc}
    2 & -1  &  0 &  0 & -1 &  0 &  0 &  0  &  0 & -1 \\
   -1 &  2  & -1 &  0 &  0 &  0 &  0 &  0  & -1 &  0 \\
    0 & -1  &  2 & -1 &  0 &  0 & -1 &  0  &  0 &  0 \\
    0 &  0  & -1 &  2 & -1 & -1 &  0 &  0  &  0 &  0 \\
   -1 &  0  &  0 & -1 &  2 &  0 &  0 & -1  &  0 &  0 \\
    0 &  0  &  0 & -1 &  0 &  2 &  0 &  0  & -1 & -1 \\
    0 &  0  & -1 &  0 &  0 &  0 &  2 & -1  &  0 & -1 \\
    0 &  0  &  0 &  0 & -1 &  0 & -1 &  2  & -1 &  0 \\
    0 & -1  &  0 &  0 &  0 & -1 &  0 & -1  &  2 &  0 \\
   -1 &  0  &  0 &  0 &  0 & -1 & -1 &  0  &  0 &  2 .
\earr
 \ear

The Dynkin diagram for this Cartan matrix could be represented by
the Petersen graph  ``a star inside a pentagon''. $P_{10}$ is a
Lorentzian KM algebra. It is a subalgebra of $E_{10}$ \cite{HPS}.

Let us present a black brane solution  for the configuration of
$10$ electric branes under consideration.  The metric
(\ref{5.2.30}) reads

\bear{5p.3.p.g} g= \left(\prod_{s =1}^{10} H_s \right)^{5/13}
\biggl\{  \left(1 -  \frac{2\mu}{R} \right)^{-1} dR \otimes dR +
R^2  d \Omega^2_{2}
\\ \nonumber
- \left( \prod_{s =1}^{10} H_s \right)^{-1} \left[ \left(1 -
\frac{2\mu}{R} \right) dt^1 \otimes dt^1  + dt^2 \otimes dt^2
\right]
\\ \nonumber
+ (H_1 H_6 H_7)^{-1}  dx^1 \otimes dx^1
\\ \nonumber
+ (H_3 H_5 H_6)^{-1}  dx^2 \otimes dx^2 + (H_5 H_7 H_9)^{-1}  dx^3
\otimes dx^3
\\ \nonumber
+  (H_1 H_4 H_8)^{-1} dx^4 \otimes dx^4 +  (H_3 H_8 H_{10})^{-1}
dx^5 \otimes dx^5
\\ \nonumber
+  (H_4 H_9 H_{10})^{-1}  dx^6 \otimes dx^6 +  (H_1 H_3 H_9)^{-1}
dx^7 \otimes  dx^7
\\ \nonumber
+  (H_2 H_6 H_8)^{-1}   dx^8 \otimes dx^8 + (H_2 H_5 H_{10})^{-1}
dx^9 \otimes  dx^9
\\ \nonumber
+ (H_2 H_4 H_7)^{-1}   dx^{10} \otimes dx^{10} \biggr\}.
\ear

The form field  is

\bear{5p.3.p.f} F^6 =  - Q_{1} R^{-2} H_1^{-2} H_2 H_5 H_{10} dR
\wedge dt^1 \wedge dt^2  \wedge dx^{1} \wedge dx^{4}\wedge dx^{7}
\\ \nonumber
- Q_{2} R^{-2} H_1 H_2^{-2} H_3 H_9 dR \wedge dt^1 \wedge dt^2
\wedge dx^{8} \wedge dx^{9}\wedge dx^{10}
\\ \nonumber
- Q_{3} R^{-2} H_2 H_3^{-2} H_4 H_7 dR \wedge dt^1 \wedge dt^2
\wedge dx^{2} \wedge dx^{5}\wedge dx^{7}
\\ \nonumber
- Q_{4} R^{-2} H_3 H_4^{-2} H_5 H_6 dR \wedge dt^1 \wedge dt^2
\wedge dx^{4} \wedge dx^{6}\wedge dx^{10}
\\ \nonumber
- Q_{5} R^{-2} H_1 H_4 H_5^{-2} H_8 dR \wedge dt^1 \wedge dt^2
\wedge dx^{2} \wedge dx^{3}\wedge dx^{9}
\\ \nonumber
- Q_{6} R^{-2} H_4  H_6^{-2} H_9 H_{10} dR \wedge dt^1 \wedge dt^2
\wedge dx^{1} \wedge dx^{2}\wedge dx^{8}
\\ \nonumber
- Q_{7} R^{-2} H_3  H_7^{-2} H_8 H_{10} dR \wedge dt^1 \wedge dt^2
\wedge dx^{1} \wedge dx^{3}\wedge dx^{10}
\\ \nonumber
- Q_{8} R^{-2} H_5 H_7 H_8^{-2} H_9 dR \wedge dt^1 \wedge dt^2
\wedge dx^{4} \wedge dx^{5}\wedge dx^{8}
\\ \nonumber
- Q_{9} R^{-2} H_2 H_6 H_8 H_9^{-2} dR \wedge dt^1 \wedge dt^2
\wedge dx^{3} \wedge dx^{6}\wedge dx^{7}
\\ \nonumber
- Q_{10} R^{-2} H_1 H_6 H_7 H_{10}^{-2} dR \wedge dt^1 \wedge dt^2
\wedge dx^{5} \wedge dx^{6}\wedge dx^{9}, \ear
 where $Q_s \neq 0$, $s = 1, \dots, 10$.

The scalar fields  reads
\beq{5p.3.ha.p} \varphi^{\alpha} = -
\frac{1}{2} \lambda_{6 \alpha} \ln \left( \prod_{s =1}^{10} H_s
\right), \eeq $\alpha = 1,2,3,4$.
Here   $H_s > 0$ obey the
equations
 \beq{5p.3.1} \frac{d}{dz} \left( \frac{(1 - 2\mu
 z)}{H_s} \frac{d}{dz} H_s \right) = 2 Q_s^2  \prod_{s' = 1}^{10}
 H_{s'}^{- A_{s s'}},
 \eeq
 with the boundary conditions
$H_{s}((2\mu)^{-1} -0) = H_{s0} \in (0, + \infty)$, and  $H_{s}(+
0) = 1$, $s = 1, ..., 10$. Here $\mu > 0$, $z = R^{-1} \in (0,
(2\mu)^{-1})$,  and  $(A_{s s'})$ is the Cartan matrix
(\ref{5a.H.A.2})  for the KM algebra $P_{10}$.

{\bf Special 1-block solution.} Now we consider a special 1-block
solution. This solution is valid if a special set of charges is
considered:  $Q_s^2 = 2 Q^2$ ($Q \neq 0$) in agreement to
(\ref{4a.3.ha.Q}) and
 \beq{4p.3.ha.b} b_s = 2 \sum_{s' =1}^{10}
 A^{ss'} = -2, \eeq
 for $s = 1, ...,  10$.
In this case the  functions $H_s$ are
 \beq{5p.4.11} H_s  =
 H^{-2}, \qquad  H =  1 + P/R, \eeq
 where $P(P + 2\mu) = 2Q^2$. As
in the previous case, we get a well-defined   solution for $P = -
\mu + \sqrt{\mu^2 + 2Q^2} > 0$ and $\mu > 0$.

The Hawking temperature in this case has the following form: $T_H
 = (1 + P/2\mu)^{10}/(8 \pi  \mu)$. It is smaller than  that in the
previous example but it also diverges  as $\mu \to + 0$. It is in
agreement  with the  singularity of the metric (\ref{5p.3.p.g}) at
$R = + 0$ for   $\mu = + 0$.

 \section{Conclusions}

We have considered several  classes of exact solutions in
multidimensional gravity with a set of scalar fields and fields of
forms  related to non-singular (e.g., hyperbolic) KM algebras.

The solutions describe composite  electromagnetic branes defined
on warped products of  Ricci-flat, or sometimes Einstein, spaces
of arbitrary dimensions and signatures. The metrics are
block-diagonal, and all scale factors, scalar fields and fields of
forms depend on points of some  manifold $M_0$. The solutions
include those depending on harmonic functions, and
spherically-symmetric black-brane solutions. Our approach is based
on the  sigma-model representation  obtained in \cite{IMC} under
rather a general assumption on intersections of composite branes
(such that the stress-energy tensor has a diagonal structure).

We have also considered  a class of  black brane configurations
with   intersection rules \cite{IMJ} governed by an invertible
generalized Cartan matrix corresponding to a certain generalized
KM Lie algebra ${\cal G}$. The ``master'' equations for moduli
functions  have polynomial solutions in the finite-dimensional
case (according   to our conjecture \cite{IMp1,IMp2,IMp3}),  while
in the infinite-dimensional case we have only a special family of
the so-called block-orthogonal solutions  corresponding to
semi-simple  non-singular  KM algebras.   Certain examples  of
black brane solutions  are presented, corresponding to the
hyperbolic KM algebras: $H_2(q,q)$ ($q > 2$), $HA_2^{(1)} =
A_2^{++}$ and the Lorentzian KM algebra  $P_{10}$.

The last two solutions (which are new) may be analyzed from the
viewpoints (i) of post-Newtonian parameters $\beta$ and $\gamma$
corresponding to the 4-dimensional section of the metric and (ii)
of  thermodynamic properties of the black branes under
consideration. These and some other tasks may be a subject of a
separate publication.

\begin{center}
{\bf Acknowledgments}
\end{center}

This work was supported in part by grant NPK-MU (PFUR), Russian
Foundation for Basic Research  (Grant  Nr. 09-02-00677-a)  and by
the  FTsP ``Nauchnie i nauchno-pedagogicheskie kadry
innovatsionnoy Rossii''  for the years 2009-2013.

\end{document}